\documentclass[conference]{IEEEtran}
\usepackage{cite}
\usepackage{amsmath,amssymb,amsfonts, autobreak, bm}
\usepackage{mathtools}
\usepackage{algorithmic}
\usepackage{graphicx}
\usepackage{textcomp}
\usepackage{xcolor}
\usepackage{relsize}
\usepackage{subcaption}
\usepackage{dblfloatfix}
\usepackage{ifthen}   

\newboolean{long}
\setboolean{long}{false}			
\newcommand{\verbose}[2]{
	\ifthenelse{\boolean{long}}{#1}{#2}
}	

\def\BibTeX{{\rm B\kern-.05em{\sc i\kern-.025em b}\kern-.08em
    T\kern-.1667em\lower.7ex\hbox{E}\kern-.125emX}}
\DeclareMathSymbol{\shortminus}{\mathbin}{AMSa}{"39}
\begin{document}
\raggedbottom
\title{Physical-Layer Authentication Using Channel State Information and Machine Learning\\
}

\author{\IEEEauthorblockN{Ken St. Germain}
\IEEEauthorblockA{\textit{Department of Electrical and Computer Engineering} \\
\textit{Naval Postgraduate School}\\
Monterey, CA \\
kenneth.stgermain@nps.edu}
\and
\IEEEauthorblockN{Frank Kragh}
\IEEEauthorblockA{\textit{Department of Electrical and Computer Engineering} \\
\textit{Naval Postgraduate School}\\
Monterey, CA \\
fekragh@nps.edu}
}

\maketitle

\begin{abstract}
Strong authentication in an interconnected wireless environment continues to be an important, but sometimes elusive goal.  Research in physical-layer authentication using channel features holds promise as a technique to improve network security for a variety of devices.  We propose the use of machine learning and measured multiple-input multiple-output communications channel information to make a decision on whether or not to authenticate a particular device.  This work analyzes the use of received channel state information from the wireless environment and demonstrates the employment of a generative adversarial neural network (GAN) trained with received channel data to authenticate a transmitting device.  We compared a variety of machine learning techniques and found that the local outlier factor (LOF) algorithm reached 100\% accuracy at lower signal to noise ratios (SNR) than other algorithms.  However, before LOF reached 100\%, we also show that the GAN was more accurate at lower SNR levels.
\end{abstract}


\section{Introduction}
\verbose{Information security \cite{noauthor_information_nodate}, the}{The} protection of integrity, confidentiality, and availability is a challenge in wireless networks.  Unlike networks with wired point-to-point connections, the broadcast nature of the wireless domain grants bona fide users and malicious actors the same access to the communication channel.  As modern and future mobile networks bring the promise of very high data rate mobile communications, they must also be secure.\verbose{  These networks will connect many different types of devices, dynamically adapt, and support minimum latency \cite{fang_security_2018}.  By leveraging multiple-input, multiple-output (MIMO) receiver/transmitter schemes, data throughput and coverage is improved while minimizing error \cite{sibille_mimo_2011}.  However, without}{  Without} appropriate security, there will be intrusions and attacks, countering the networks' benefits.\verbose{  There is no lack of reporting on computer network attacks, and continued justification in improving against a myriad of threats at the federal, state, and private levels \cite{ferdinando_terabyte_nodate, konkel_pentagon_2018, gallagher_hackers_nodate, noauthor_hackers_2019}.}{}

\verbose{Knowing the identity of the person or device you are communicating with is a vital part of secure communications.  Without a robust method to ensure authenticity, a fake message can be sent or a transmitted message can be altered.  The typical authentication scenario involves two entities, Alice and Bob, authenticating with each other.  That is, Alice is communicating to Bob and providing some sort of information to prove to Bob that Alice is indeed Alice and not someone else.  Bob, in turn, will also prove himself to Alice.  A third entity, Eve, can also communicate with Alice and Bob and eavesdrop on their conversation.  Strong authentication ensures that Eve can't pose as either Alice or Bob.  Authentication is often employed at various levels of the Open Systems Interconnected (OSI) model:  application-layer \cite{dizdarevic_survey_2019}, transport-layer \cite{fantacci_analysis_2007}, network-layer  \cite{xing_security_2008}, media access control-layer \cite{aime_dependability_2007}, and physical-layer  \cite{baracca_physical_2012}.  While employing authentication at different layers may enhance wireless security, the overhead remains a debt to be paid in computational complexity and latency \cite{apostolopoulos_securing_2000, zou_survey_2016}.}{} 

\verbose{Motivated by the desire to ensure strong authentication on a wide variety of devices with disparate applications, we explore the use of physical-layer authentication.  Authentication is commonly and successfully accomplished using key-based cryptography, however as networks grow and become more complex, key distribution and management may not scale without causing undue user delays \cite{fantacci_analysis_2007, wang_physical-layer_2016}.  Unfortunately, key-based systems still have vulnerabilities that a motivated adversary can exploit, and in some instances very little security \cite{fluhrer_weaknesses_2001} is gained at the expense of increased complication or worse quality of service \cite{menasce_qos_2002}.  Another drawback for cryptographic systems is the challenge of key distribution and management in a decentralized, dynamic, and heterogeneous network \cite{liu_asymmetric_2009, wang_physical-layer_2016}.  Depending on the application, there are many networked systems that forego any protective measures at all.  For example, because of issues such as energy needs, processing capability, and storage requirements, many Internet of Things (IoT) devices do not support strong security mechanisms yet they connect to a greater network becoming an attractive ingress point for malicious actors \cite{meneghello_iot:_2019}.  However, when authentication is accomplished at the physical-layer, the burden on higher layers in the OSI model is reduced, saving computational burden on the device and preventing an undue expenditure of resources.}{}

Our ultimate intent is to classify a transmitter as either trusted or untrusted and use that information to make a decision on whether or not to authenticate.  Machine learning tools have shown success at producing solutions for complex problems where data is consistently changing and difficult to characterize with simpler techniques \cite{geron_hands-machine_2017}.  To this end, we require unique transmitter features realized in the physical-layer environment.

\verbose{The literature proposes two broad categories to distinguish legitimate from illegitimate devices at the physical-layer \cite{zeng_non-cryptographic_2010, wang_physical-layer_2016} without the use of a pre-shared secret, cryptography, user-provided credentials, or higher OSI-layer processing.}{The literature proposes two broad categories to distinguish legitimate from illegitimate devices at the physical-layer.}  The first relies on unique imperfections of the transmitter hardware that manifest as radio frequency (RF) fingerprints or signatures \cite{brik_wireless_2008, gungor_basic_2016, polak_identifying_2011}.\verbose{  Based on manufacturing processes and designs, the transmitted signal will be uniquely distorted from device to device, even if only slightly.}{}The second method leverages the stochastic nature of the wireless channel to take advantage of multi-path fading environments.  The temporally and spatially-unique impulse or frequency response can be used to identify the transmitter \cite{tugnait_wireless_2013, yingbin_liang_secure_2008, al_khanbashi_real_2013}.

Our proposed method is based on research using the second category.  The effects of the multipath channel can be described in the channel state information (CSI) matrix.\verbose{  The elements of the CSI matrix are attributes of the fading channel and are therefore unique to the pairwise position of the receiver and transmitter in line-of-sight (LOS) and non-line-of-sight (NLOS) multipath environments.  While there is merit in using the RF fingerprinting method in the first category, these characteristics are observable by a malicious actor and can be spoofed.  Contrast with using the channel-based approach, an adversary cannot directly measure the CSI between two entities and create a transmission to mimic a legitimate signal.  In dynamic conditions with a mobile transmitter, receiver, and/or significant reflective or absorbing objects, the CSI is also time-variant.}{}The focus of this paper is on the static case, and using a technique as described by \cite{xiao_physical-layer_2008} can be adopted to account for scenarios where motion is expected to change the CSI.

In this paper we use a generative adversarial network (GAN) to determine if a transmitter should be authenticated or denied access.

The contributions of this paper are:
\begin{itemize}
	\item We retain both magnitude and phase information to make an authentication decision.
	\item We introduce analysis and simulation illustrating how the received CSI matrix elements and measurement error can be used for physical-layer authentication.
	
	\verbose{\item We propose a generative model that creates fake CSI matrix samples closely matching the characteristics of legitimate CSI matrix samples from a trusted source.}{}
	\verbose{\item We propose a discriminative model that processes both legitimate samples from a trusted source and faked samples created by the generative model.}{}
	\verbose{\item The models are validated by simulation.  The generative and discriminative models were trained and tested on datasets modeling multipath NLOS Rayleigh channels.}{}
	\item There are two novel contributions in this work: 
	\subitem (1) In Section \ref{authentication}, a hypothesis test for physical-layer authentication using all elements in a CSI matrix and the respective receiver measurement error on those elements.
	\subitem (2) In Section \ref{simulation}, the use of a GAN model to accurately use MIMO CSI as a basis of authentication at the physical-layer without requiring a survey of CSI at various positions.  
	\item Our proposal discards the generative model at the conclusion of training and retains the discriminative model.  The fully-trained discriminative model, having learned from a generative model that creates indistinguishably realistic CSI samples, is particularly suited to make authentication decisions.
	
\end{itemize}

This paper discusses previous work in physical-layer authentication using machine learning in Section~\ref{background}.  Section~\ref{authentication} provides the concept for authentication using CSI and introduces a method to accomplish this.  Next we present the system model for the GAN in Section~\ref{system}.  The development of the GAN and simulation results is shown in Section~\ref{simulation}.  Finally, we summarize our observations and discuss future work in Section~\ref{conclusion}.  With respect to notation, unless otherwise addressed, vectors are indicated with bold lower-case letters, and matrices are bold upper-case letters.

\section{Background and Related Work}\label{background}

This section introduces the channel model and discusses related work with machine learning, including the application of GANs.

\verbose{By taking advantage of the randomness and uniqueness inherent in the RF communication channel,  physical-layer information provided by the channel can be used to conduct authentication \cite{xiao_fingerprints_2007, liu_physical_2017}.}{}  
\subsection{Channel model}
The nature of the wireless medium affects the transmitted signal as it propagates to the receiver.  The narrowband model of the wireless channel\verbose{ \cite{verdu_spectral_2002, choi_coding_2019}}{}is given by 
\begin{equation}\label{eq:channel}
{\textbf{\textit{y}}=\textbf{\textit{H}}\textbf{\textit{x}}+\textbf{\textit{n}}}
\end{equation}
where \textit{\textbf{y}} is the received signal, \textit{\textbf{x}} is the transmitted signal, \textit{\textbf{H}} is the time-varying CSI or channel response, and \textit{\textbf{n}} is the noise vector.  \textit{\textbf{H}} is an ${ N \times M }$ matrix of circularly symmetric complex-valued Gaussian random variables representing multiple channel conditions such as multi-path fading and the use of multiple antennas \cite{yu_models_2002}.  The number of transmitter antennas is $ M $ and the number of receiver antennas is $ N $.  Each complex element within \textit{\textbf{H}} is composed of real and imaginary zero-mean independent Gaussian random variables with identical variance and with Rayleigh distributed magnitude for NLOS scenarios.  Jakes' uniform scattering model \cite{jakes_microwave_1995} states that antennas spatially separated more than two carrier wavelengths from each other will observe sufficiently independent fading channels due to rapid decorrelation of the signal envelope among receivers.

\verbose{  Fig.~\ref{fig:channel} illustrates the wireless channel model and CSI matrix.  
\begin{figure}
	\centering     
	\includegraphics[height=3in]{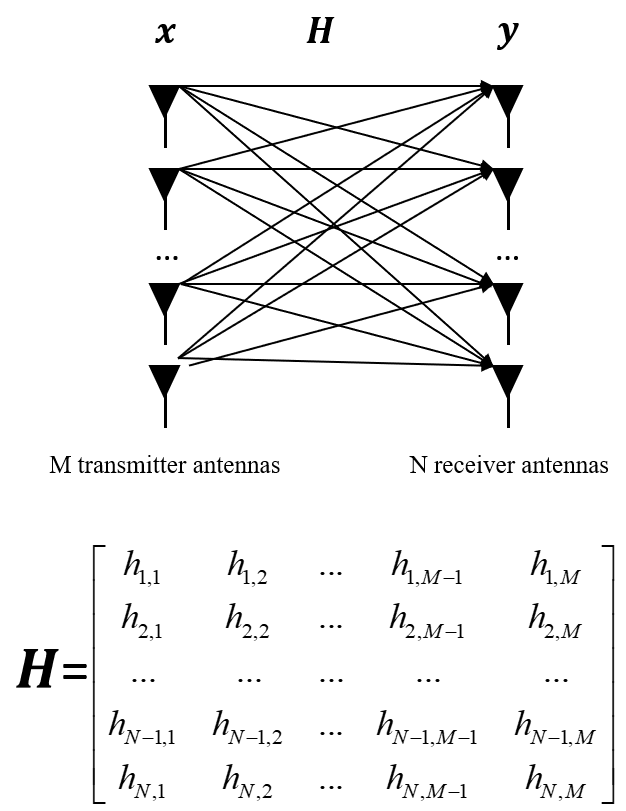}
	\small{MIMO narrowband channel model and CSI matrix}
	\label{fig:channel}
\end{figure}

By taking advantage of the spatially-unique properties of the CSI, many researchers have leveraged machine learning and deep neural networks in particular to successfully resolve localization challenges.}{}  

\subsection{CSI authentication through machine learning}
Many works have been conducted with machine learning and location information to make an authentication decision based on CSI \cite{liu_authenticating_2018, liao_novel_2019, wang_deep_2019, liao_deep-learning-based_2019, liao_security_2019,   Abyaneh_deep_2018}.  Support vector machines (SVM), K-means clustering, and an algorithm to determine temporal correlation of CSI magnitude were used by Liu et al. in \cite{liu_authenticating_2018}.  In \cite{liao_novel_2019}, Liao et al. used a convolutional neural network (CNN) to identify the valid users from attackers, based on a training dataset that included different path delays among the various users.  Wang et al. \cite{wang_deep_2019} also used this methodology, but included a recurrent neural network (RNN), a hybrid CNN-RNN architecture, and a skip-layer CNN.  In \cite{liao_deep-learning-based_2019} and \cite{liao_security_2019}, Liao et al. again used a variety of neural networks to identify valid users and spoofers using pre-built training sets created by obtaining CSI from legitimate and illegal sensor node locations.  In \cite{Abyaneh_deep_2018}, Abyaneh et al. also used a CNN trained on pre-mapped transmitter locations in order to grant authentication, however, they also included the complex value of the CSI element as input to their neural network.

To implement many proposals, researchers pre-processed collected raw CSI and used the magnitude of the CSI elements, $|h_{n,m}|$, in their implementations.  Additionally, those that use neural networks used training data that required advance knowledge of CSI between a monitor and possible legitimate and illegitimate transmitters positions.  Our proposal retains the complex elements of the CSI to retain a rich set of features and our methodology does not need a priori CSI samples from an attacker's position while we make use of unsupervised machine learning to reach an authentication decision.

Supervised learning through neural networks have helped to make advancements in a variety of fields and  require training data to respond well when tested.  If the task is for classification, training examples enable the network to accurately label subsequent samples.  However, there are machine learning algorithms that are designed to detect outliers or anomalies.

\subsection{Machine learning for outlier and anomaly detection}

There is broad use for detection systems that identify samples with features outside of the expected.  A one-class machine learning problem trains on data from a single class, and during testing the tool will provide an output recognizing new samples as either inliers of the statistical distribution or outliers.

Local outlier factor (LOF) from Breunig et al. in \cite{Breunig_LOF_2000} quantifies the relative degree of isolation of a sample to its neighbors.  For samples with density similar to that of the local cluster, the LOF value approaches 1.  Inlier samples with higher density than their neighbors have a LOF less than 1, and outliers have a LOF greater than 1.

Isolation forest (iForest) from the works of Liu et al. in \cite{Liu_Isolation_2008, Liu_Isolation-Based_2012} does not use density measurements to detect anomalies, but randomly selects a threshold between maximum and minimum values of a randomly selected feature.  The iForest algorithm takes advantage of the defining characteristics of an anomaly:  there are few of them, and they have features unlike normal samples.  

The one-class support vector machine (OC-SVM) as presented by Schölkopf et al. in \cite{Scholkopf_Estimating_2001} is an extension to the traditional SVM technique, but uses unsupervised learning and unlabeled training data.  The goal of the OC-SVM algorithm is to determine the smallest region containing the training data.  The algorithm returns a value of 1 if a new sample falls within this region, and a -1 is returned for samples outside the threshold.

A large advantage in implementing these one-class machine learning tools is the speed at which they can be trained and tested \cite{Liu_Isolation-Based_2012}.  Unfortunately, the performance of these tools may not scale as the dimensionality of the task increases \cite{Pedregosa_Scikit-learn_2011}.

\subsection{GAN research in communication systems}

Goodfellow et al. proposed the novel concept of a GAN in \cite{goodfellow_generative_2014-1}.  Composed of two artificial neural network models called the discriminator and the generator, the GAN framework trains these models as they compete against each other and mutually improve.  While GANs have successfully contributed to many areas that rely on image processing such as single image super-resolution \cite{ledig_photo-realistic_2017}, medical radiology \cite{armanious_medgan:_2019}, facial recognition \cite{bao_cvae-gan:_2017}, etc., there have also been breakthroughs by applying GANs to investigations in the wireless communication.  

{To minimize symbol error, O'Shea et al. \cite{oshea_physical_2018} used a GAN to determine the optimal modulation scheme in a given channel, showing how GANs can allow for adaptation to the RF environment.  In an adversarial situation such as jamming and spoofing, Roy et al. \cite{roy_rfal:_2019} proposed a GAN-based method to determine legitimate from illegitimate transmitters based on the imbalance of in-phase and quadrature components of a symbol constellation.  The amplitude-feature deep convolutional GAN was used by Li et al. \cite{li_af-dcgan:_2019} to reduce the effort and increase the accuracy in creating a CSI-based fingerprint database for a Wi-Fi localization system.}


\section{Authentication with CSI}\label{authentication}
A receiver continues to authenticate a transmitter if the received CSI varies less than a threshold applied to the received CSI from previous transmissions.  \verbose{This requires some method of initial authentication, such as the use of cryptographic methods or physical-layer authentication using RF fingerprinting from transmitter imperfections.}{}During initial authentication,\verbose{}{by such means as cryptography or RF fingerprinting,}the receiver makes CSI measurements of the channel and stores that information for future authentication.

During channel measurement,\verbose{even in a stable static environment,}{}the receiver imparts noise to the received signal, resulting in variation to the measured CSI elements.  This error, $\boldsymbol\epsilon$, is modeled as an additive complex zero-mean Gaussian process, $\mathcal{C}\mathcal{N}(0, \boldsymbol\Sigma_{\epsilon}) $, where the covariance of the sample mean is $ \boldsymbol\Sigma_{\overline{\epsilon}} = \boldsymbol\Sigma_{\epsilon}/s $  for $ s $ samples during the measurement.  Therefore, the ${k}$th CSI measured by the receiver, $ \hat{\textbf{\textit{H}}}_k $, is given as   
\begin{equation}\label{eq:received_CSI}
\hat{\textbf{\textit{H}}}_k=\textbf{\textit{H}}+\bm{\epsilon}_{{k}}
\qquad {k} = 1, 2, \dots , s
\end{equation}
where $ \textbf{\textit{H}} $ is the true CSI from (\ref{eq:channel}) and $ \bm{\epsilon}_{\mathit{k}} $ is a complex ${ N \times M }$ matrix with independent identically distributed elements.\verbose{  Since $ \bm{\epsilon}_{k} $ is zero-mean, $ \textbf{\textit{H}} $, can be estimated with a variety of techniques including least squares estimation, minimum mean-square error estimation, and through successive measurements and element-wise averaging of $ \hat{\textbf{\textit{H}}}_{k} $ for $ k = \{1,2,\dots,s\} $ as demonstrated in \cite{chapre_csi-mimo_2014}.}{}  

A threshold is then applied to each CSI element, $ \textit{h}_{n,m} $, where the transmitter is authenticated if the distance from every received element, $ \hat{\textit{h}}_{n,m,k} $ from $ \hat{\textbf{\textit{H}}}_{k} $ for $k>s$,  to the estimated element, $ \textit{h}_{n,m} $ from $ \textbf{\textit{H}} $, is less than or equal to a threshold, $ {\textit{z}}_{n,m} $, based on the average eigenvalue, $ \lambda_{ave} $, from the covariance matrix  $ \boldsymbol{\Sigma}_{\overline{\epsilon}_{n,m}} $.  To simplify the notation, we will consider $ {\textit{z}}_{n,m} $ the same value $z$ for all $n$ and $m$ terms, however in practice, $z$ could vary among CSI elements.  The numbered sequential transmission count is represented by $ k $. Following the hypothesis testing in \cite{xiao_using_2008-1}, we have the null hypothesis, $ \mathcal{H}_{0} $, to authenticate, and the alternative hypothesis, $ \mathcal{H}_{1} $, to deny authentication
\begin{equation}\label{eq:hypothesis}
\begin{split}
\mathcal{H}_{0}: &(\text{Re}(\hat{\textit{h}}_{n,m,k}) - \text{Re}(\textit{h}_{n,m}))^2 \\
&\qquad+ (\text{Im}(\hat{\textit{h}}_{n,m,k}) - \text{Im}(\textit{h}_{n,m}))^2 \leq {\textit{z}}^2 \quad \boldsymbol{\forall} n,m\\\\
\mathcal{H}_{1}: &(\text{Re}(\hat{\textit{h}}_{n,m,k}) - \text{Re}(\textit{h}_{n,m}))^2 \\
&\qquad+ (\text{Im}(\hat{\textit{h}}_{n,m,k}) - \text{Im}(\textit{h}_{n,m}))^2 > {\textit{z}}^2 \quad \boldsymbol{\exists} n,m
\end{split}
\end{equation}	
where $\text{Re}(\mathord{\cdot}) $ and $\text{Im}(\mathord{\cdot}) $ return the real and imaginary parts of the CSI matrix elements, respectively and $ \textit{z} $, is a tunable parameter that can be adjusted to suit the requirements of the system.  To minimize false positives, $ {\textit{z}} $ can be set to a relatively small value such as $ \lambda_{ave}^{\frac{1}{2}} $, and to minimize false negatives, $ \textit{z} $ can be expanded to a greater value, such as $ 6 \lambda_{ave}^{\frac{1}{2}} $.\verbose{ As evident in the equation forms for $ \mathcal{H}_{0} $ and $ \mathcal{H}_{1} $, the regions for authentication are circles of radius $ z $ centered on the coordinates provided by the CSI elements, $ h_{n,m} $.}{}In order to successfully authenticate, all elements in $ \textbf{\textit{H}} $ and $ \textit{z} $ must jointly achieve the $ \mathcal{H}_{0} $ result.  

\begin{figure}
	\includegraphics[height=2.5in]{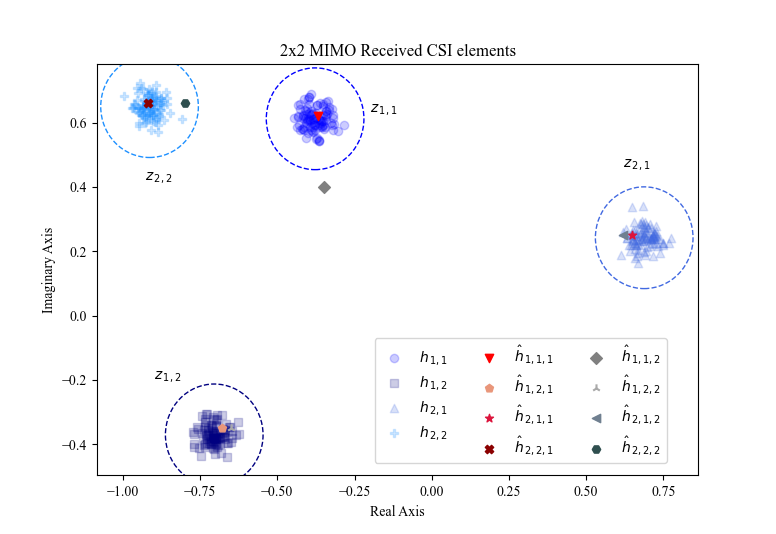}
	\caption{Measured 2x2 MIMO CSI elements with receiver noise}
	\label{fig:CSIerror}
\end{figure}
As an example, consider Fig.~\ref{fig:CSIerror} illustrating the measured CSI elements for the ${ 2 \times 2 }$ MIMO case.  The markers in shades of blue indicate measured samples gathered during the initial authentication.  Comparing the red-shaded CSI elements from $ \hat{\textit{h}}_{n,m,1} $ and gray-shaded elements in $ \hat{\textit{h}}_{n,m,2} $, the desired outcome is that $ \hat{\textit{h}}_{n,m,1} $ will authenticate, but $ \hat{\textit{h}}_{n,m,2} $ will not authenticate due to $ \hat{\textit{h}}_{1,1,2} $ likely being outside of the error tolerance for $ z_{1,1} $, where for this case $ z = 5 \lambda_{ave}^{\frac{1}{2}} $ for all $n$$m$ terms.

Given an ${ N \times M }$ array of normally distributed random variables, we can determine the probability of one transmitter being accidentally authenticated as another based on the error tolerance for the first transmitter, $ \textit{z} $.  Let $ a_{n,m} $ be the real part and $ b_{n,m} $ be the imaginary part of the complex value for the true CSI element, $ h_{n,m} =a_{n,m} + j  b_{n,m} $.  Both $ a_{n,m} $ and $ b_{n,m} $ are independent Gaussian random variables with variance $ \sigma^{2} /2 $.\verbose{The joint probability distribution function (PDF) for $ a_{n,m} $ and $ b_{n,m} $ is 
\begin{equation}\label{eq:joint_pdf}
\begin{gathered}
f(a_{n,m},b_{n,m}) = \frac{\exp{\left(-\frac {a_{n,m}^{2}+b_{n,m}^{2}}{ \sigma^2}\right)}}{2\pi \sqrt{|\boldsymbol\Sigma_{a_{n,m},b_{n,m}}|}}  \\
\boldsymbol\mu_{a_{n,m},b_{n,m}} = \begin{pmatrix} 0 \\ 0 \end{pmatrix}, \qquad
\boldsymbol\Sigma_{a_{n,m},b_{n,m}} = \begin{pmatrix} \sigma^{2}/2 & 0 \\
0  & \sigma^{2}/2 \end{pmatrix}
\end{gathered}
\end{equation}
}{}To this variable, we add the result of the receiver noise, $ \boldsymbol\epsilon $.  The real and imaginary parts of $ \epsilon_{n,m} $ are zero-mean independent Gaussian distributed random variables each with sample mean covariance $  \Sigma_{\overline{\epsilon}_{n,m}} $. 

For a transmitter to be authenticated, $ \mathcal{H}_{0} $ must be satisfied for every CSI element, $ \textit{h}_{n,m} $.  Given $ \textit{h}_{n,m} = a_{n,m} + j b_{n,m} $, we can determine the probability that another transmitter will be authenticated.  Let $ z = 5 \lambda_{ave}^{\frac{1}{2}} $ where $ \lambda_{ave} $ is the average eigenvalue from the receiver noise covariance matrix, $ \boldsymbol\Sigma_{\overline{\epsilon}_{n,m}} $ and $ u $ and $ v $ be the respective real and imaginary parts of the CSI from another transmitter.  The probability of $ u + jv $ resulting in $ \mathcal{H}_{0} $ for $ h_{n,m} $ is
\begin{equation}\label{eq:probability_of_overlap}
\begin{split}
P([u+jv] \in \mathcal{D}_{n,m}) = \iint_{\mathcal{D}_{n,m}}
\frac{
	\exp{\left(
		-\frac{
			u^{2}+v^{2}}
		 {\sigma^{2} }\right) } }{2\pi\sqrt{|\boldsymbol\Sigma_{u,v}|} }\,du\,dv 
\\
\noalign{where,}
\mathcal{D}_{n,m} = \{(u,v)\mid \left(u-a_{n,m}\right)^2+\left(v-b_{n,m}\right)^2 \leq z^2\}\\\\
\boldsymbol\mu_{u,v} = \begin{pmatrix} 0 \\ 0 \end{pmatrix}, \qquad
\boldsymbol\Sigma_{u,v} = \begin{pmatrix} \sigma^{2}/2 & 0 \\
0  & \sigma^{2}/2 \end{pmatrix}
\end{split}
\raisetag{5\baselineskip}
\end{equation}
With independent $ u $ and $ v $, (\ref{eq:probability_of_overlap}) can be evaluated using ${ P(X \cap Y) = P(Y|X) \cdot P(X) }$, where $ P(X) $ is the probability that ${ {a_{n,m}-z} \leq \enspace u \enspace \leq {a_{n,m}+z} }$, and $ P(Y|X) $ is the probability that ${ {b_{n,m}\shortminus\sqrt{z^2\shortminus(u\shortminus a_{n,m})^2}} \leq  v  \leq {b_{n,m}+\sqrt{z^2\shortminus(u\shortminus a_{n,m})^2}} }$.  Therefore, 
\begin{equation}\label{eq:probability}
\begin{split}
\begin{aligned}
P([u+jv]& \in \mathcal{D}_{n,m}) = \left(Q(A)-Q(B)\right) \cdot \left(Q(C)-Q(D)\right) \\
\noalign{where, }
&A = \frac{a_{n,m}-z}{\sigma} \quad B = \frac{a_{n,m}+z}{\sigma} \\
&C = \frac{b_{n,m}-\sqrt{z^2-\left(u-a_{n,m}\right)^2}}{\sigma} \\
&D = \frac{b_{n,m}+\sqrt{z^2-\left(u-a_{n,m}\right)^2}}{\sigma} \\
\noalign{and the $ Q(\cdot) $ function is}
&Q(x) = \int_x^{+\infty} \frac{1}{\sqrt{2\pi}}\exp\left(-\frac{t^2}{2}\right)dt
\end{aligned}
\end{split}
\raisetag{7\baselineskip}
\end{equation}

\begin{figure}[b]
	\centering
	\includegraphics[height=2.2in]{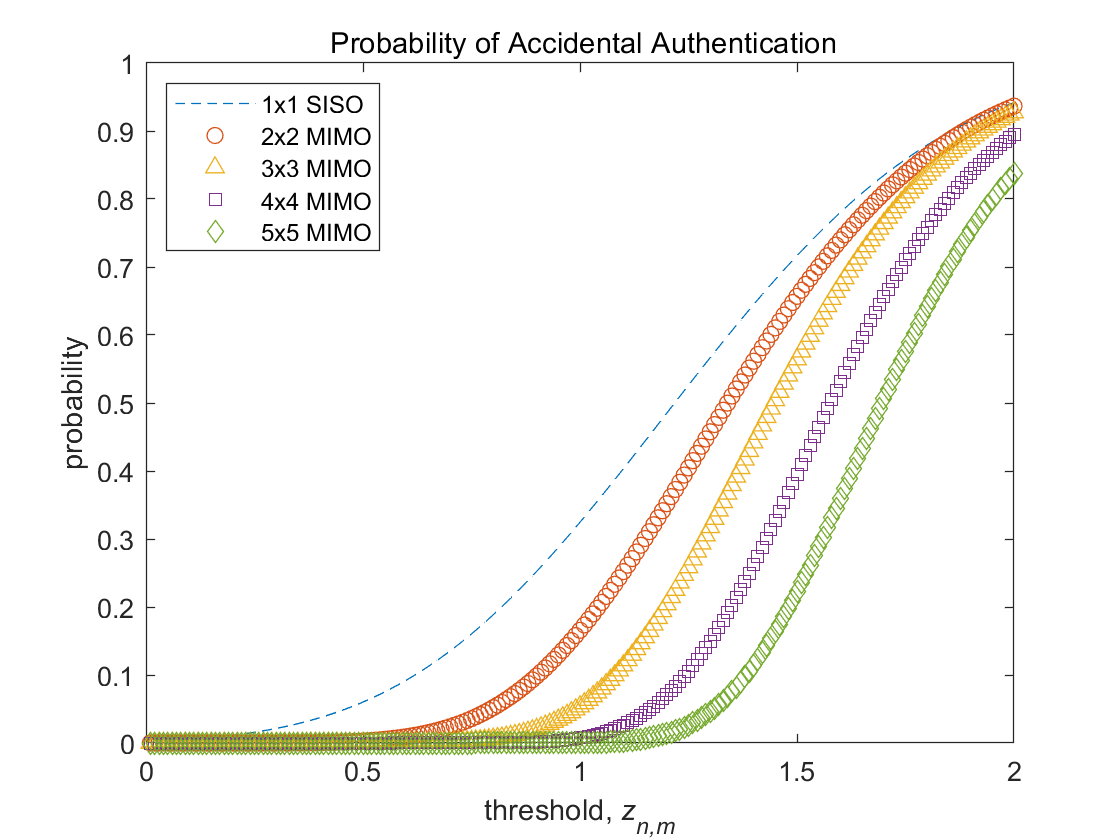}
	\caption{Probability of authentication for various MIMO configurations and thresholds}
	\label{fig:prob_auth}
\end{figure}

The transmitter must satisfy $ \mathcal{H}_{0} $ for every CSI element.  The probability for authentication in a MIMO channel with $ M $ transmit antennas and $ N $ receive antennas is then

\begin{equation}\label{eq:product_auth}
\begin{gathered}
\prod_{m=1}^{M} \prod_{n=1}^{N} P([u_{n,m}+jv_{n,m}] \in \mathcal{D}_{n,m})\\
\mathcal{D}_{n,m} = \{(u_{n,m},v_{n,m})\mid \left(u_{n,m}-a_{n,m}\right)^2
\\ \qquad \qquad+\left(v_{n,m}-b_{n,m}\right)^2
 \leq z^2\}\\
\end{gathered}
\end{equation}
Simulating (\ref{eq:probability}) and (\ref{eq:product_auth}) with $ a_{n,m}, b_{n,m}, u,$ and $ v $  all distributed as $ \mathcal{N}(0,0.5) $, Fig.~\ref{fig:prob_auth} illustrates how unlikely an accidental authentication will be as the number of antenna elements of the receiver and transmitter are increased and the threshold is reduced.

To implement this authentication scheme and determine which hypothesis $ \hat{\textit{h}}_{n,m,k} $ satisfies, we require advance knowledge of the noise power our receiver imparts to $\textbf{\textit{H}} $ to determine $z$ and that may change over time and be different among devices.  Instead, we will allow a neural network to implicitly determine the threshold and perform the authentication decision.  We created a GAN that is trained on authentic samples from a dataset and samples produced by a generative model.  The discriminative model then learned the characteristics of $ \textit{h}_{n,m} $ and $ \boldsymbol\Sigma_{\overline{\epsilon}_{n,m}} $.  Following training, two testing datasets validated the performance of the discriminative model to accurately distinguish $ \textbf{\textit{H}} $ between trusted and untrusted transmitters.
  
\section{System Model}\label{system}
\begin{figure}[b]
	\centering     
	\includegraphics[height=1.35in]{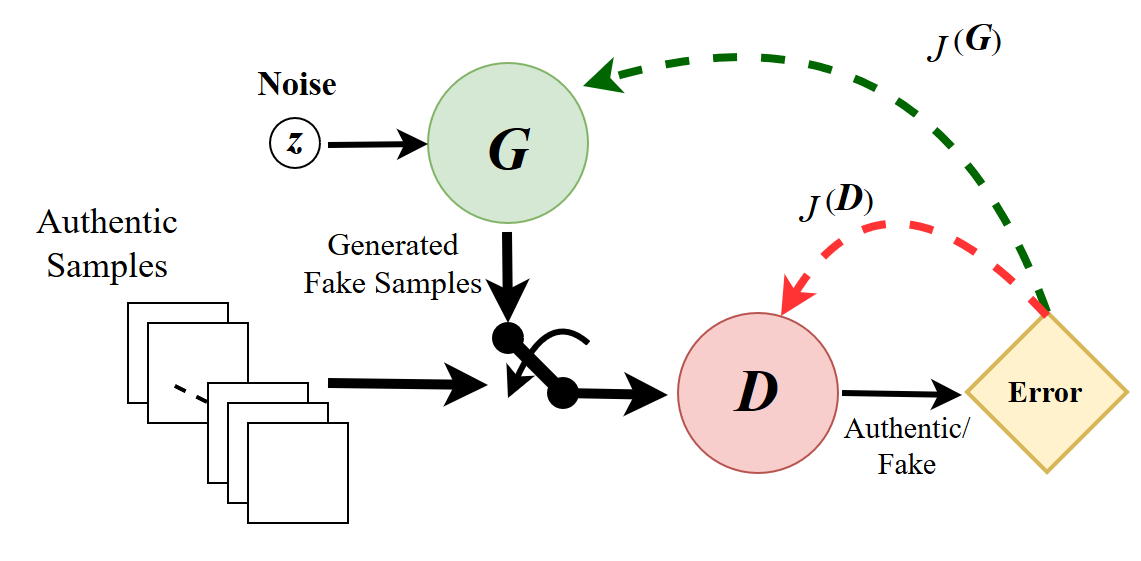}
	\caption{Training a generative adversarial network}
	\label{fig:gan}
\end{figure}
We consider a wireless MIMO communications channel with trusted users and untrusted users, some of the latter group who are malicious adversaries.  The adversaries have resources available to change their antenna characteristics, transmitter RF path timing, output power, and/or present reflectors between themselves and the receiver.  Thus, they are able to change their CSI as measured by the receiver.  To defeat this scenario, the discriminative model at the receiver is adversarially trained by a generative model that creates authentic looking CSI samples.  \verbose{By training with increasingly high quality spoofed samples, the discriminative network learns the features of transmitters that should be authenticated and the features of those that should not be authenticated.

\subsection{GAN Architecture}}{}


During training, the discriminative model, $\mathcal{D}$, receives authentic samples from the training data or fake samples generated by the generative model, $\mathcal{G}$.  The generative model creates fake samples based on a function from random variable input, $\mathit{z}$, and the parameters in $\mathcal{G}$. The discriminative model then assigns a probability from zero to one based on whether the sample is fake (0.0) or authentic (1.0).  Fig.~\ref{fig:gan} shows a functional depiction of a GAN in training, where $\mathit{J^{(D)}}$ and $\mathit{J^{(G)}}$ are the loss functions for the discriminative model and the generative model, respectively.

The adversarial competition in the GAN is a minimax game where the discriminative model attempts to correctly label training samples from a distribution produced by CSI matrix elements, $p_{data}(h_{n,m})$, and fake training samples created by the generator.  The discriminative model is trained to maximize the probability of assigning the correct label, while the generative model is trained to minimize the same probability.  The value function that describes this relationships from the original work by Goodfellow \cite{goodfellow_generative_2014-1} is given by 
\begin{equation}\label{eq:minimax}
\begin{split}
\min_G \max_D V(D, G)&= \mathbb{E}_{x\sim p_{data}(x)}[\log D(x)] 
\\
&\qquad + \mathbb{E}_{z\sim p_{z}(z)}[\log(1 - D(G(z)))]
\end{split}
\end{equation}
where $\mathit{D(x)}$ is the probability that $\mathit{x}$ came from the data distribution $\mathit{p_{data}(x)}$ containing authentic training samples, and $\mathit{D(G(z))}$ is the estimate of the probability that the discriminator incorrectly identifies the fake instance as authentic. The generator network attempts to maximize $\mathit{D(G(z))}$, while the discriminator network tries to minimize it.  The generator creates samples, $\mathit{G(z)}$, based on the parameter values in $\mathit{G}$ and the random values $z$ provided to the generator consistent with $\mathit{p_{z}(z)}$.

As each entity adversarially trains each other, they learn to improve their individual performance.  When the discriminative model correctly identifies fake samples created by the generative model, the generative network will update its parameter weights through backpropagation to make more realistic samples.  Likewise, the discriminative model will update its parameter weights when it incorrectly identifies real or fake samples.  The results of this training are a generator neural network adept at creating data that closely mimics training data and a discriminator neural network that can identify all but the best fakes.  
\verbose{
\subsection{Discriminative Model}

The discriminator estimates the probability that a sample came from the training data, rather than the generator.  When training begins, the discriminator won't know $\mathit{p_{data}(x)}$, so the accuracy of correctly assigning authentic and fake samples will be near 0.5.  The accuracy will increase with more iterations of samples and backpropagation as the authentic data distribution is learned until the generator network creates samples such that the fake sample distribution, $\mathit{p_{g}(z)}$ optimally matches $\mathit{p_{data}(x)}$.  At this point, the accuracy of correctly assigning authentic and fake samples will return to 0.5 since for the optimal discriminator $D^*$, and fixed generator, $G$, $\mathit{D}^*_\mathit{G}(x) = \frac{\mathit{p_{data}(x)}}{\mathit{p_{data}(x)}+\mathit{p_{z}(z)}}$.  When $\mathit{p_{data}(x)} = \mathit{p_{z}(z)}$, $\mathit{D}^*_\mathit{G}(x) = 0.5$ \cite{goodfellow_generative_2014}.

\subsection{Generative Model}

Without having direct access to $\mathit{p_{data}(x)}$, the generator attempts to capture this distribution through feedback based on the probabilities the discriminator assigns to generated fake samples \cite{goodfellow_generative_2014}.  The weights of the generator network are updated via the loss function $\mathit{J^{(G)}}$ so that the generator will create better samples.}{}

\section{Simulation}\label{simulation}

The GAN processed a single subcarrier in a MIMO ${ 4 \times 4 }$ configuration.  Therefore, the discriminative model has 16 complex inputs and 1 real output, while the generative model has 1 real input and 16 complex outputs.  The inputs for the discriminative model and the outputs for the generative model represent the complex elements in the CSI matrix.  

\subsection{GAN development}

\begin{table}[]
	\centering
	\caption{GAN architecture}
	\begin{tabular}{llll}
		\multicolumn{2}{l}{Discriminator:}                  &                                  &                                \\ \cline{2-4} 
		& \multicolumn{1}{c|}{Layer}                       & \multicolumn{1}{c|}{output size} & \multicolumn{1}{c}{activation} \\ \cline{2-4} 
		& \multicolumn{1}{l|}{Input 1: $x\sim p_{data}(x_{1,1})$}  & \multicolumn{1}{l|}{2}           &                                \\
		& \multicolumn{1}{l|}{Input 2: $x\sim p_{data}(x_{1,2})$}  & \multicolumn{1}{l|}{2}           &                                \\
		& \multicolumn{1}{l|}{\vdots}                         & \multicolumn{1}{l|}{\vdots}         &                                \\
		& \multicolumn{1}{l|}{Input 16: $x\sim p_{data}(x_{4,4})$} & \multicolumn{1}{l|}{2}           &                                \\
		& \multicolumn{1}{l|}{Concatenated}                 & \multicolumn{1}{l|}{32}          &                                \\
		& \multicolumn{1}{l|}{Fully connected}             & \multicolumn{1}{l|}{64}          & LeakyReLU (alpha = 0.3)        \\
		& \multicolumn{1}{l|}{Dropout = 0.2}             & \multicolumn{1}{l|}{}          &         \\
		& \multicolumn{1}{l|}{Fully connected}             & \multicolumn{1}{l|}{32}          & LeakyReLU (alpha = 0.3)        \\
		& \multicolumn{1}{l|}{Dropout = 0.2}             & \multicolumn{1}{l|}{}          &         \\
		& \multicolumn{1}{l|}{Output}                      & \multicolumn{1}{l|}{1}           & Sigmoid                        \\ \cline{2-4} 
		&                                                  &                                  &                                \\
		\multicolumn{2}{l}{Generator:}                      &                                  &                                \\ \cline{2-4} 
		& \multicolumn{1}{l|}{Layer}                       & \multicolumn{1}{l|}{output size} & activation                     \\ \cline{2-4} 
		& \multicolumn{1}{l|}{Input:  $z\sim p_{z}(z)$}      & \multicolumn{1}{l|}{5}           &                                \\
		& \multicolumn{1}{l|}{Fully connected}             & \multicolumn{1}{l|}{16}          & LeakyReLU (alpha = 0.3)        \\
		& \multicolumn{1}{l|}{Fully connected}             & \multicolumn{1}{l|}{32}          & LeakyReLU (alpha = 0.3)        \\
		& \multicolumn{1}{l|}{Fully connected}             & \multicolumn{1}{l|}{64}          & tanh                           \\
		& \multicolumn{1}{l|}{Output 1}                    & \multicolumn{1}{l|}{2}           & linear                         \\
		& \multicolumn{1}{l|}{Output 2}                    & \multicolumn{1}{l|}{2}           & linear                         \\
		& \multicolumn{1}{l|}{\vdots}                         & \multicolumn{1}{l|}{\vdots}         & \vdots                            \\
		& \multicolumn{1}{l|}{Output 16}                   & \multicolumn{1}{l|}{2}           & linear                         \\ \cline{2-4} 
	\end{tabular}
	\label{tab:gan_full}
\end{table}

The GAN is implemented using the Python programming language, Keras \cite{chollet_et_al_keras_2015} front-end, and Tensorflow \cite{abadi_et_al_tensorflow_2015} back-end.  Additionally, Numpy, Pandas, and Matplotlib Python libraries were used.  The overall GAN design is summarized in Table~\ref{tab:gan_full}, with a total of 19,425 parameters.  The file size of the discriminator network was 104 KB.  

The discriminator network, $\mathcal{D}$, has 16 inputs of size 2 merged into one \textit{concatenated} layer.  Each input has 2 nodes to accommodate the real and imaginary parts of the complex CSI matrix element.  Two additional fully connected layers of size 64 and 32 with \textit{LeakyReLU} activations (\textit{alpha} = 0.3) follow.  Both of these hidden layers use \textit{Dropout} of 0.2 to prevent overfitting.  The output layer of size 1 is fully connected and uses a \textit{sigmoid} activation to provide values [0.0,~1.0].  The learning rate for $\mathcal{D}$ was 0.0003 using the \textit{Adam} \cite{kingma_adam:_2017} optimizer.

The generator network, $\mathcal{G}$, has a single input with 5 nodes fully connected to the first hidden layer of size 16.  Two additional hidden layers of sizes 32 and 64 are again fully connected using \textit{LeakyReLU} (\textit{alpha} = 0.3) and \textit{tanh} activations respectively.  Finally, 16 output layers of size~2 are connected using \textit{linear} activations.  The learning rate for $\mathcal{G}$ was 0.0009 using the \textit{Adam} optimizer.
\subsection{Datasets}
A master dataset was created by adding measurement error in the form of AWGN across a range of signal to noise ratio (SNR) levels to a single ${4 \times 4}$ CSI matrix composed of 16 circularly symmetric Gaussian complex values with zero mean, and unit variance, $\mathcal{CN}(0,1)$.  The SNR values ranged from 0~dB to 30~dB in steps of 2~dB, and 1,000 samples were created at each SNR level.  Each sample is a ${ 4 \times 4 }$ complex matrix.

Splitting evenly across SNR levels, the training dataset uses 70\% of the master dataset samples, reserving 30\% for the testing dataset.  Two testing datasets were created, each consisting of 700 samples for each SNR value.  In addition to the valid samples taken from the master dataset, 400 more testing samples were created to simulate two different operating scenarios. 

The first testing dataset replicated the accidental authentication case.  There are six transmitters, one of which should be authenticated.  The 300 samples taken from the master dataset represent the transmitter that should be authenticated.  For the remaining transmitters, five new ${4 \times 4}$ CSI matrices with elements taken from $\mathcal{CN}(0,1)$ were created.  To each matrix, AWGN at varying SNR values was added to produce 80 samples at each SNR value.  These samples were then added to the accidental authentication dataset, resulting in 300 legitimate samples and 400 illegitimate samples.  We will refer to this dataset as the accidental authentication test dataset.  

\begin{figure}[]
	\begin{tabular}[b]{c}
		\includegraphics[width=.23\textwidth]{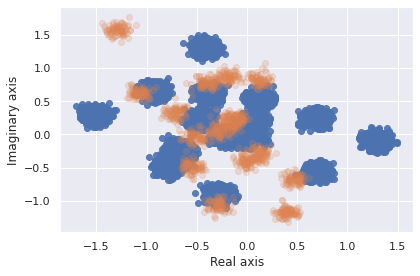} 
		\includegraphics[width=.23\textwidth]{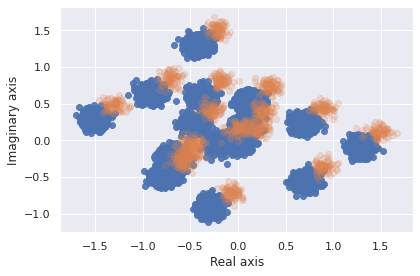} \\
	\end{tabular} 
	
	\hspace{0.9 in} \small (a) 
	\hspace{1.5 in}
	\small (b) 
	
	\caption{Samples from the (a) accidental authentication and  (b) nefarious users test datasets at 20 dB.}
	\label{fig:samples}
\end{figure}

The second testing dataset emulated five nefarious users attempting to authenticate by matching the CSI matrix of a single legitimate transmitter.  If by some unlikely method, an adversary were able to know the channel characteristics between two legitimately authenticated transmitters, such as described by Shi et al. in \cite{shi_generative_2019}, the adversary may also have the resources necessary to spoof their transmitted CSI to appear as another transmitter's received CSI.  As before, 300 samples from the master dataset for each of the 16 SNR levels ranging from 0~dB to 30~dB in steps of 2~dB are the legitimate samples.  To complete this dataset, five different complex number offsets were added to the original legitimate CSI matrix.  Samples were created as previously described by the addition of AWGN again resulting in 300 legitimate samples and 400 illegitimate samples.  We will refer to this dataset as the nefarious users test dataset.

Samples from each dataset are shown in Fig.~\ref{fig:samples}, where blue dots correspond to the 16 clusters of legitimate samples, and the orange circles represent illegitimate samples that should not be authenticated.  Note that only one example from the illegitimate test group is shown for each dataset.  Training was restricted to a maximum of 50 epochs in mini-batches of 64 samples.  

\subsection{Results}
\begin{figure}[]
	\centering
	\begin{tabular}[b]{c}
		\includegraphics[width=.23\textwidth]{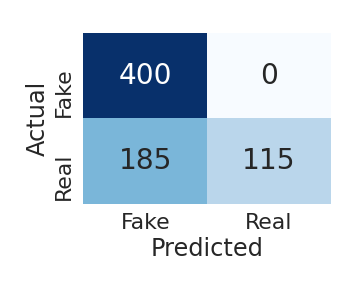} 
		\includegraphics[width=.23\textwidth]{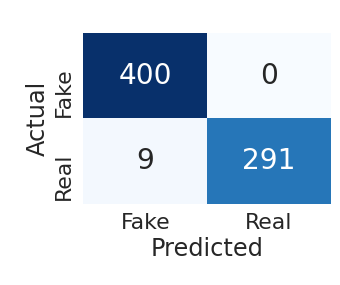} \\
		\small (a) SNR = 0 dB \hspace{0.7in}
		\small (b) SNR = 4 dB\\
	\end{tabular} 
	\begin{tabular}[b]{c}
	\end{tabular}
	
	\begin{tabular}[b]{c}
		\includegraphics[width=.23\textwidth]{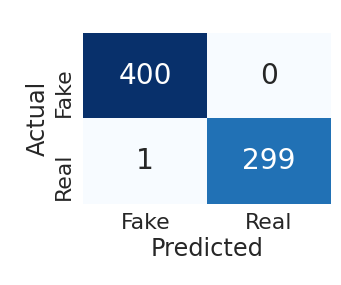} 
		\includegraphics[width=.23\textwidth]{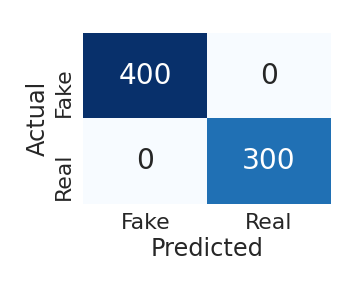} \\
		\small (c) SNR = 8 dB \hspace{0.7in}
		\small (d) SNR = 10 dB
	\end{tabular}
	
	\caption{GAN discriminator performance against accidental authentication test dataset with SNR levels at (a)~0~dB, (b)~4~dB, (c)~8~dB, (d)~12~dB. }
	\label{fig:accidental}
\end{figure}
The performance of the discriminative model indicates the viability of using a GAN for physical-layer authentication using CSI.  To illustrate the performance, we use confusion matrices to show how the discriminator assigns samples from the test datasets.  The horizontal axis is the predicted result from the discriminator on whether or not authentication should be granted.  If the discriminator assesses the sample to be legitimate, the sample is tallied in the ``Real" column.  If the discriminator's result predict that the sample is illegitimate and should not be authenticated, the sample is allocated to the ``Fake" column.  The vertical axis is the ground truth of the sample from the dataset, where legitimate samples are assigned to the ``Real" row, and illegitimate samples are shown in the ``Fake" row.

Against the accidental authentication testing dataset, the discriminator achieves 100\% accuracy for SNR greater than or equal to 10 dB.  As shown in Fig.~\ref{fig:accidental}, for SNR less than 10 dB, the discriminator makes errors in correctly identifying legitimate samples but never allows illegitimate samples to be authenticated.

\begin{figure}[]
	\centering
	\begin{tabular}[b]{c}
		\includegraphics[width=.23\textwidth]{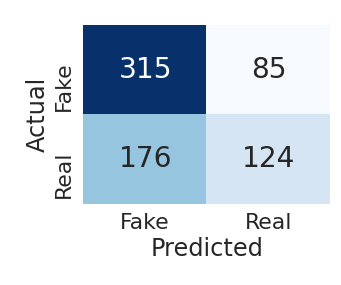} 
		\includegraphics[width=.23\textwidth]{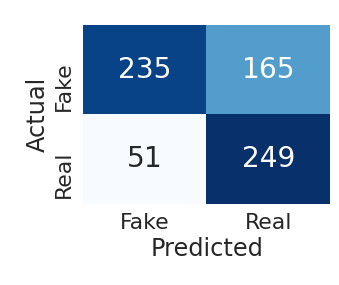} \\
		\small (a) SNR = 0 dB \hspace{0.7in}
		\small (b) SNR = 4 dB\\
	\end{tabular} 
	\begin{tabular}[b]{c}
	\end{tabular}
	
	\begin{tabular}[b]{c}
		\includegraphics[width=.23\textwidth]{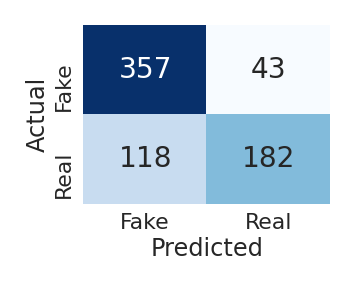} 
		\includegraphics[width=.23\textwidth]{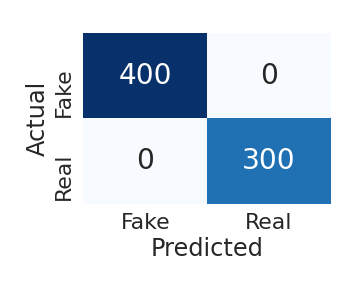} \\
		\small (c) SNR = 8 dB \hspace{0.7in}
		\small (d) SNR = 20 dB
	\end{tabular}

	\caption{GAN discriminator performance against nefarious users test dataset with SNR levels at (a)~0~dB, (b)~4~dB, (c)~8~dB, (d)~20~dB. }
	\label{fig:nefarious}
\end{figure}

For the nefarious user testing dataset, the discriminator doesn't achieve 100\% until the SNR reaches 20 dB.  As shown in  Fig.~\ref{fig:nefarious}, in addition to mischaracterizing legitimate samples, the discriminator allows some number of illegitimate samples to be authenticated, since there are samples counted in the ``Fake" row and ``Real" column.

In order to compare the GAN-trained discriminator's performance, we apply the same datasets to the hypothesis test in (\ref{eq:hypothesis}), and to three machine learning algorithms.  Because the master dataset does not contain illegitimate samples, the techniques we use are limited to one-class, novelty, or anomaly detection algorithms.  We employ LOF, iForest, and OC-SVM.  These techniques are available and were implemented from the Scikit-learn project \cite{Pedregosa_Scikit-learn_2011}.

\begin{figure}[]
	\centering     
	\includegraphics[width=.47\textwidth]{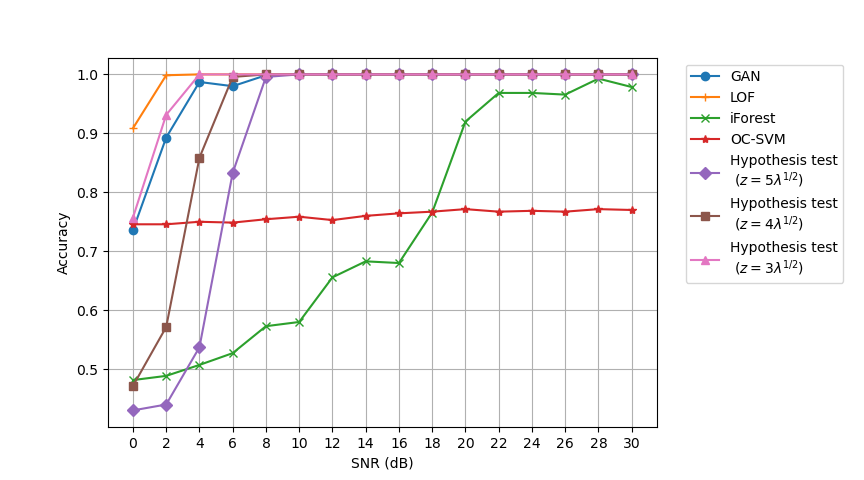}
	\caption{Accuracy vs SNR for accidental authentication dataset}
	\label{fig:accidental_fig}
\end{figure}

%
%

For the accidental authentication test dataset, Fig.~\ref{fig:accidental_fig} shows the accuracy performance for the machine learning techniques.  We see that the LOF algorithm is superior for all SNR levels and that the GAN and hypothesis test with ${z = 3\lambda^{\frac{1}{2}}}$ are very close in performance.

\begin{figure}[]
	\centering     
	\includegraphics[width=.47\textwidth]{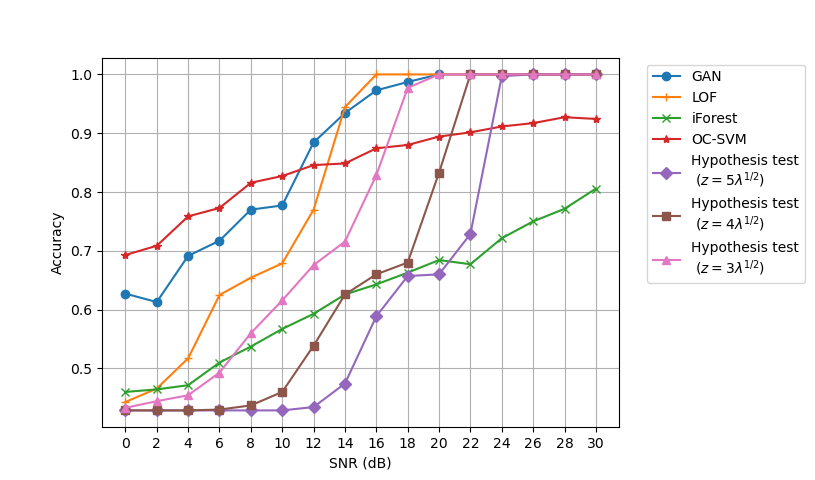}
	\caption{Accuracy performance vs SNR for nefarious users dataset}
	\label{fig:resourceful_fig}
\end{figure}

Fig.~\ref{fig:resourceful_fig} illustrates the accuracy of the machine learning techniques against the nefarious users test dataset.  While LOF is the first algorithm to reach 100\% accuracy, its performance at low SNR values against this dataset is outmatched by the GAN.  We noted that all techniques including the GAN, as we've seen in Fig.~\ref{fig:nefarious}, incorrectly categorize illegitimate samples as ``Real" for low SNR against the nefarious users who are able to closely match the CSI of legitimate transmitters.  

Adjustment of hyperparameters and training methodology for the GAN can be applied to improve results.  For example, by increasing the mini-batch size, we saw an improvement in identifing the illegitimate samples.  Unfortunately, this also caused a reduction in the number of legitimate samples identified.  Changes to the original GAN architecture in this work is left for future refinement.

\verbose{After each epoch of training, the discriminator was saved, and subsequently tested against both datasets.  From the 22\textsuperscript{nd} through 30\textsuperscript{th} epochs, both test results successfully authenticated the 250 legitimate samples, and determined the remaining 1,250 were not legitimate.}{}

\verbose{At the onset of training, the discriminator accuracy increases from approximately 0.5 until the generator loss function motivated more accurate samples.  Training continued with alternating rising and falling loss functions for both models in the GAN until they stabilized at roughly 0.7.  At that point, the ability of the discriminator to discern legitimate samples from generated samples was 50\%, indicating that the generated samples very closely matched the legitimate samples.}{}  
	
\verbose{Following testing on both datasets, the discriminator output was examined.  The confidence level in the discriminator to assign legitimate samples to as ``Real" was comparable for both datasets.  However, the discriminator was more confident in assigning the ``Fake" label to the testing dataset using the accidental case, and less confident (but still correct) in assigning the "Fake" label in the five nefarious user case.}{}


\section{Conclusion and Future Work}\label{conclusion}

We showed how CSI could be used as a method to provide physical-layer authentication.  Our analysis illustrated that the probability of accidentally authenticating other transmitters decreases as the number of receive and transmit antennas are increased and a threshold value is judiciously applied.  We then developed a GAN trained on a dataset of CSI matrices to perform physical-layer authentication in an adversarial environment.  After training less than 50 epochs, the discriminator reached 100\% accuracy on two seperate testing datasets, implicitly determining appropriate thresholds for received CSI matrix elements.  Across all levels of AWGN, the LOF algorithm was best at reaching 100\% accuracy.  

This paper demonstrated how physical-layer authentication can be accomplished in a flat fading single subchannel environment.  By applying this concept to multiple subchannels, there is an opportunity for obtaining robust multi-channel characteristics that can be used to identify a transmitter for authentication.  Furthermore, the hyperparameters of the GAN should be optimized and reevaluated on additional training and test datasets to demonstrate effectiveness in a variety of wireless environments.

%

\bibliographystyle{IEEEtran}
\bibliography{StGermain_Asilomar_2020}

\begin{thebibliography}{10}
\providecommand{\url}[1]{#1}
\csname url@samestyle\endcsname
\providecommand{\newblock}{\relax}
\providecommand{\bibinfo}[2]{#2}
\providecommand{\BIBentrySTDinterwordspacing}{\spaceskip=0pt\relax}
\providecommand{\BIBentryALTinterwordstretchfactor}{4}
\providecommand{\BIBentryALTinterwordspacing}{\spaceskip=\fontdimen2\font plus
\BIBentryALTinterwordstretchfactor\fontdimen3\font minus
  \fontdimen4\font\relax}
\providecommand{\BIBforeignlanguage}[2]{{%
\expandafter\ifx\csname l@#1\endcsname\relax
\typeout{** WARNING: IEEEtran.bst: No hyphenation pattern has been}%
\typeout{** loaded for the language `#1'. Using the pattern for}%
\typeout{** the default language instead.}%
\else
\language=\csname l@#1\endcsname
\fi
#2}}
\providecommand{\BIBdecl}{\relax}
\BIBdecl

\bibitem{geron_hands-machine_2017}
\BIBentryALTinterwordspacing
A.~Géron, \emph{\BIBforeignlanguage{en}{Hands-on machine learning with
  {Scikit}-{Learn} and {TensorFlow} : concepts, tools, and techniques to
  build intelligent systems}}, first edition~ed.\hskip 1em plus 0.5em minus
  0.4em\relax Sebastopol, CA: O'Reilly Media, 2017. [Online]. Available:
  \url{http://shop.oreilly.com/product/0636920052289.do}
\BIBentrySTDinterwordspacing

\bibitem{brik_wireless_2008}
V.~Brik, S.~Banerjee, M.~Gruteser, and S.~Oh,
  ``\BIBforeignlanguage{en}{Wireless device identification with radiometric
  signatures},'' in \emph{\BIBforeignlanguage{en}{Proceedings of the 14th {ACM}
  international conference on {Mobile} computing and networking - {MobiCom}
  '08}}.\hskip 1em plus 0.5em minus 0.4em\relax San Francisco, California, USA:
  ACM Press, 2008, p. 116.

\bibitem{gungor_basic_2016}
O.~Gungor and C.~E. Koksal, ``On the {Basic} {Limits} of
  {RF}-{Fingerprint}-{Based} {Authentication},'' \emph{IEEE Transactions on
  Information Theory}, vol.~62, no.~8, pp. 4523--4543, Aug. 2016.

\bibitem{polak_identifying_2011}
A.~C. Polak, S.~Dolatshahi, and D.~L. Goeckel, ``Identifying {Wireless} {Users}
  via {Transmitter} {Imperfections},'' \emph{IEEE Journal on Selected Areas in
  Communications}, vol.~29, no.~7, pp. 1469--1479, Aug. 2011.

\bibitem{tugnait_wireless_2013}
J.~K. Tugnait, ``Wireless {User} {Authentication} via {Comparison} of {Power}
  {Spectral} {Densities},'' \emph{IEEE Journal on Selected Areas in
  Communications}, vol.~31, no.~9, pp. 1791--1802, Sep. 2013.

\bibitem{yingbin_liang_secure_2008}
{Yingbin Liang}, H.~Poor, and S.~Shamai, ``\BIBforeignlanguage{en}{Secure
  {Communication} {Over} {Fading} {Channels}},''
  \emph{\BIBforeignlanguage{en}{IEEE Transactions on Information Theory}},
  vol.~54, no.~6, pp. 2470--2492, Jun. 2008.

\bibitem{al_khanbashi_real_2013}
N.~Al~Khanbashi, N.~Al~Sindi, S.~Al-Araji, N.~Ali, Z.~Chaloupka, V.~Yenamandra,
  and J.~Aweya, ``Real time evaluation of {RF} fingerprints in wireless {LAN}
  localization systems,'' in \emph{2013 10th {Workshop} on {Positioning},
  {Navigation} and {Communication} ({WPNC})}, Mar. 2013, pp. 1--6.

\bibitem{xiao_physical-layer_2008}
L.~Xiao, L.~Greenstein, N.~Mandayam, and W.~Trappe, ``A {Physical}-{Layer}
  {Technique} to {Enhance} {Authentication} for {Mobile} {Terminals},'' in
  \emph{2008 {IEEE} {International} {Conference} on {Communications}}, May
  2008, pp. 1520--1524, iSSN: 1550-3607, 1938-1883.

\bibitem{yu_models_2002}
\BIBentryALTinterwordspacing
K.~Yu and B.~Ottersten, ``\BIBforeignlanguage{en}{Models for {MIMO} propagation
  channels: a review},'' \emph{\BIBforeignlanguage{en}{Wireless Communications
  and Mobile Computing}}, vol.~2, no.~7, pp. 653--666, 2002. [Online].
  Available: \url{https://onlinelibrary.wiley.com/doi/abs/10.1002/wcm.78}
\BIBentrySTDinterwordspacing

\bibitem{jakes_microwave_1995}
W.~C. Jakes, Ed., \emph{\BIBforeignlanguage{en}{Microwave mobile
  communications}}, nachdr.~ed., ser. An {IEEE} {Press} classic reissue.\hskip
  1em plus 0.5em minus 0.4em\relax New York, NY: IEEE Press [u.a.], 1995, oCLC:
  249569885.

\bibitem{liu_authenticating_2018}
H.~Liu, Y.~Wang, J.~Liu, J.~Yang, Y.~Chen, and H.~V. Poor, ``Authenticating
  {Users} {Through} {Fine}-{Grained} {Channel} {Information},'' \emph{IEEE
  Transactions on Mobile Computing}, vol.~17, no.~2, pp. 251--264, Feb. 2018,
  conference Name: IEEE Transactions on Mobile Computing.

\bibitem{liao_novel_2019}
R.~Liao, H.~Wen, F.~Pan, H.~Song, A.~Xu, and Y.~Jiang, ``A {Novel} {Physical}
  {Layer} {Authentication} {Method} with {Convolutional} {Neural} {Network},''
  in \emph{2019 {IEEE} {International} {Conference} on {Artificial}
  {Intelligence} and {Computer} {Applications} ({ICAICA})}, Mar. 2019, pp.
  231--235, iSSN: null.

\bibitem{wang_deep_2019}
Q.~Wang, H.~Li, D.~Zhao, Z.~Chen, S.~Ye, and J.~Cai, ``Deep {Neural} {Networks}
  for {CSI}-{Based} {Authentication},'' \emph{IEEE Access}, vol.~7, pp.
  123\,026--123\,034, 2019, conference Name: IEEE Access.

\bibitem{liao_deep-learning-based_2019}
\BIBentryALTinterwordspacing
R.-F. Liao, H.~Wen, J.~Wu, F.~Pan, A.~Xu, Y.~Jiang, F.~Xie, and M.~Cao,
  ``Deep-{Learning}-{Based} {Physical} {Layer} {Authentication} for
  {Industrial} {Wireless} {Sensor} {Networks},'' \emph{Sensors (Basel,
  Switzerland)}, vol.~19, no.~11, May 2019. [Online]. Available:
  \url{https://www.ncbi.nlm.nih.gov/pmc/articles/PMC6603790/}
\BIBentrySTDinterwordspacing

\bibitem{liao_security_2019}
R.~Liao, H.~Wen, J.~Wu, F.~Pan, A.~Xu, H.~Song, F.~Xie, Y.~Jiang, and M.~Cao,
  ``Security {Enhancement} for {Mobile} {Edge} {Computing} {Through} {Physical}
  {Layer} {Authentication},'' \emph{IEEE Access}, vol.~7, pp.
  116\,390--116\,401, 2019.

\bibitem{Abyaneh_deep_2018}
\BIBentryALTinterwordspacing
A.~Y. Abyaneh, A.~H.~G. Foumani, and V.~Pourahmadi, ``Deep {Neural} {Networks}
  {Meet} {CSI}-{Based} {Authentication},'' \emph{arXiv:1812.04715 [cs, eess,
  stat]}, Nov. 2018, arXiv: 1812.04715. [Online]. Available:
  \url{http://arxiv.org/abs/1812.04715}
\BIBentrySTDinterwordspacing

\bibitem{Breunig_LOF_2000}
\BIBentryALTinterwordspacing
M.~M. Breunig, H.-P. Kriegel, R.~T. Ng, and J.~Sander, ``{LOF}: identifying
  density-based local outliers,'' \emph{ACM SIGMOD Record}, vol.~29, no.~2, pp.
  93--104, May 2000. [Online]. Available:
  \url{http://doi.org/10.1145/335191.335388}
\BIBentrySTDinterwordspacing

\bibitem{Liu_Isolation_2008}
F.~T. Liu, K.~M. Ting, and Z.-H. Zhou, ``Isolation {Forest},'' in \emph{2008
  {Eighth} {IEEE} {International} {Conference} on {Data} {Mining}}, Dec. 2008,
  pp. 413--422, iSSN: 2374-8486.

\bibitem{Liu_Isolation-Based_2012}
\BIBentryALTinterwordspacing
------, ``Isolation-{Based} {Anomaly} {Detection},'' \emph{ACM Transactions on
  Knowledge Discovery from Data}, vol.~6, no.~1, pp. 3:1--3:39, Mar. 2012.
  [Online]. Available: \url{http://doi.org/10.1145/2133360.2133363}
\BIBentrySTDinterwordspacing

\bibitem{Scholkopf_Estimating_2001}
\BIBentryALTinterwordspacing
B.~Schölkopf, J.~C. Platt, J.~Shawe-Taylor, A.~J. Smola, and R.~C. Williamson,
  ``\BIBforeignlanguage{en}{Estimating the {Support} of a {High}-{Dimensional}
  {Distribution}},'' \emph{\BIBforeignlanguage{en}{Neural Computation}},
  vol.~13, no.~7, pp. 1443--1471, Jul. 2001. [Online]. Available:
  \url{https://www.mitpressjournals.org/doi/abs/10.1162/089976601750264965}
\BIBentrySTDinterwordspacing

\bibitem{Pedregosa_Scikit-learn_2011}
\BIBentryALTinterwordspacing
F.~Pedregosa, G.~Varoquaux, A.~Gramfort, V.~Michel, B.~Thirion, O.~Grisel,
  M.~Blondel, P.~Prettenhofer, R.~Weiss, V.~Dubourg, J.~Vanderplas, A.~Passos,
  D.~Cournapeau, M.~Brucher, M.~Perrot, and Ã.~Duchesnay, ``Scikit-learn:
  {Machine} {Learning} in {Python},'' \emph{Journal of Machine Learning
  Research}, vol.~12, no.~85, pp. 2825--2830, 2011. [Online]. Available:
  \url{http://jmlr.org/papers/v12/pedregosa11a.html}
\BIBentrySTDinterwordspacing

\bibitem{goodfellow_generative_2014-1}
\BIBentryALTinterwordspacing
I.~J. Goodfellow, J.~Pouget-Abadie, M.~Mirza, B.~Xu, D.~Warde-Farley, S.~Ozair,
  A.~Courville, and Y.~Bengio, ``Generative {Adversarial} {Nets},'' in
  \emph{Proceedings of the 27th {International} {Conference} on {Neural}
  {Information} {Processing} {Systems} - {Volume} 2}, ser. {NIPS}'14.\hskip 1em
  plus 0.5em minus 0.4em\relax Cambridge, MA, USA: MIT Press, 2014, pp.
  2672--2680, event-place: Montreal, Canada. [Online]. Available:
  \url{http://dl.acm.org/citation.cfm?id=2969033.2969125}
\BIBentrySTDinterwordspacing

\bibitem{ledig_photo-realistic_2017}
C.~Ledig, L.~Theis, F.~Huszár, J.~Caballero, A.~Cunningham, A.~Acosta,
  A.~Aitken, A.~Tejani, J.~Totz, Z.~Wang, and W.~Shi, ``Photo-{Realistic}
  {Single} {Image} {Super}-{Resolution} {Using} a {Generative} {Adversarial}
  {Network},'' in \emph{2017 {IEEE} {Conference} on {Computer} {Vision} and
  {Pattern} {Recognition} ({CVPR})}, Jul. 2017, pp. 105--114, iSSN: 1063-6919.

\bibitem{armanious_medgan:_2019}
\BIBentryALTinterwordspacing
K.~Armanious, C.~Jiang, M.~Fischer, T.~Küstner, T.~Hepp, K.~Nikolaou,
  S.~Gatidis, and B.~Yang, ``\BIBforeignlanguage{en}{{MedGAN}: {Medical}
  {Image} {Translation} using {GANs}},''
  \emph{\BIBforeignlanguage{en}{Computerized Medical Imaging and Graphics}}, p.
  101684, Nov. 2019. [Online]. Available:
  \url{http://www.sciencedirect.com/science/article/pii/S0895611119300990}
\BIBentrySTDinterwordspacing

\bibitem{bao_cvae-gan:_2017}
J.~Bao, D.~Chen, F.~Wen, H.~Li, and G.~Hua, ``{CVAE}-{GAN}: {Fine}-{Grained}
  {Image} {Generation} through {Asymmetric} {Training},'' in \emph{2017 {IEEE}
  {International} {Conference} on {Computer} {Vision} ({ICCV})}, Oct. 2017, pp.
  2764--2773, iSSN: 2380-7504.

\bibitem{oshea_physical_2018}
T.~J. O'Shea, T.~Roy, N.~West, and B.~C. Hilburn, ``Physical {Layer}
  {Communications} {System} {Design} {Over}-the-{Air} {Using} {Adversarial}
  {Networks},'' in \emph{2018 26th {European} {Signal} {Processing}
  {Conference} ({EUSIPCO})}, Sep. 2018, pp. 529--532.

\bibitem{roy_rfal:_2019}
D.~Roy, T.~Mukherjee, M.~Chatterjee, E.~Blasch, and E.~Pasiliao, ``{RFAL}:
  {Adversarial} {Learning} for {RF} {Transmitter} {Identification} and
  {Classification},'' \emph{IEEE Transactions on Cognitive Communications and
  Networking}, pp. 1--1, 2019.

\bibitem{li_af-dcgan:_2019}
Q.~Li, H.~Qu, Z.~Liu, N.~Zhou, W.~Sun, S.~Sigg, and J.~Li, ``{AF}-{DCGAN}:
  {Amplitude} {Feature} {Deep} {Convolutional} {GAN} for {Fingerprint}
  {Construction} in {Indoor} {Localization} {Systems},'' \emph{IEEE
  Transactions on Emerging Topics in Computational Intelligence}, pp. 1--13,
  2019.

\bibitem{xiao_using_2008-1}
L.~Xiao, L.~J. Greenstein, N.~B. Mandayam, and W.~Trappe, ``Using the physical
  layer for wireless authentication in time-variant channels,'' \emph{IEEE
  Transactions on Wireless Communications}, vol.~7, no.~7, pp. 2571--2579, Jul.
  2008.

\bibitem{chollet_et_al_keras_2015}
\BIBentryALTinterwordspacing
F.~Chollet, et~al., \emph{Keras}, 2015. [Online]. Available:
  \url{https://keras.io}
\BIBentrySTDinterwordspacing

\bibitem{abadi_et_al_tensorflow_2015}
\BIBentryALTinterwordspacing
M.~Abadi, et~al., ``{TensorFlow}: {Large}-{Scale} {Machine} {Learning} on
  {Heterogeneous} {Systems},'' 2015. [Online]. Available:
  \url{https://www.tensorflow.org/}
\BIBentrySTDinterwordspacing

\bibitem{kingma_adam:_2017}
\BIBentryALTinterwordspacing
D.~P. Kingma and J.~Ba, ``Adam: {A} {Method} for {Stochastic} {Optimization},''
  \emph{arXiv:1412.6980 [cs]}, Jan. 2017, arXiv: 1412.6980. [Online].
  Available: \url{http://arxiv.org/abs/1412.6980}
\BIBentrySTDinterwordspacing

\bibitem{shi_generative_2019}
\BIBentryALTinterwordspacing
Y.~Shi, K.~Davaslioglu, and Y.~E. Sagduyu, ``\BIBforeignlanguage{en}{Generative
  {Adversarial} {Network} for {Wireless} {Signal} {Spoofing}},'' in
  \emph{\BIBforeignlanguage{en}{Proceedings of the {ACM} {Workshop} on
  {Wireless} {Security} and {Machine} {Learning} - {WiseML} 2019}}.\hskip 1em
  plus 0.5em minus 0.4em\relax Miami, FL, USA: ACM Press, 2019, pp. 55--60.
  [Online]. Available:
  \url{http://dl.acm.org/citation.cfm?doid=3324921.3329695}
\BIBentrySTDinterwordspacing

\end{thebibliography}

\end{document}